# LUMINOSITY DETERMINATION AT THE TEVATRON*

V. Papadimitriou[#], Fermilab, Batavia, IL 60510, U.S.A.

*Abstract*

In this paper we discuss the luminosity determination at the Tevatron. We discuss luminosity measurements by the machine as well as by using the luminosity detectors of the CDF and D0 experiments. We discuss the uncertainties of the measurements, the effort to maximize the initial and integrated luminosity, the challenges and the lessons learned.

## INTRODUCTION

Luminosity measurements are an absolutely necessary component of any experimental beam colliding program since they provide the frequency of the interactions and the needed normalization for the physics processes under study. Luminosity measurements also allow for the monitoring of the performance of the accelerator and for implementation of beam parameter adjustments as needed for optimized performance. We describe here absolute luminosity measurements by the machine based on the measurement of beam parameters, and real time, relative luminosity measurements performed by CDF and D0 which are then normalized to the inclusive, inelastic proton-antiproton cross section.

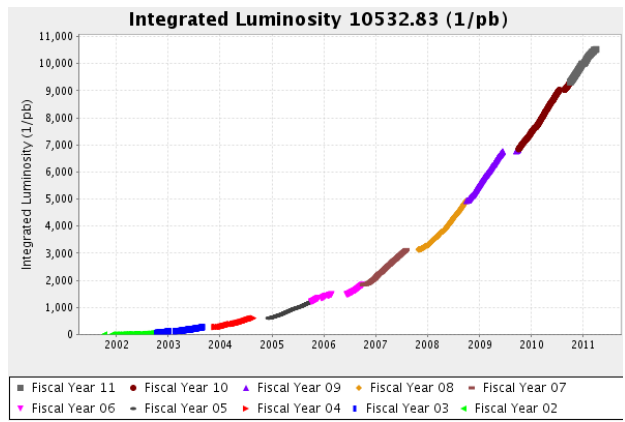

Figure 1: Tevatron integrated luminosity in Run II (averaged over the CDF and D0 experiments) as a function of time.

## TEVATRON PERFORMANCE

In 2010 the Tevatron Collider celebrated the 25[th] anniversary of proton-antiproton collisions that took place for the first time on October 17, 1985. The Tevatron, a proton-antiproton collider, has delivered 110 pb$^{-1}$ per experiment to the CDF and D0 experiments in Run I (1992 – 1996) at a center of mass energy of 1.8 TeV. In Run I, six proton bunches were colliding with six antiproton bunches with a 3.5 μs spacing between collisions. Since the beginning of Run II and by the end of March 2011 the Tevatron has in addition delivered 10.5 fb$^{-1}$ per experiment at a center of mass energy of 1.96 TeV (see Fig. 1). In Run II, thirty six proton bunches are colliding with thirty six antiproton bunches with a 396 ns spacing between collisions. The proton and antiproton beams in the Tevatron share a common vacuum pipe. Their paths are controlled by electrostatic separators which keep the beams apart around most of the Tevatron ring and bring them to collision at the CDF and D0 Interaction Points (IPs).

On April 16, 2010 there was set an initial luminosity record of 4.024 x 10$^{32}$ cm$^{-2}$s$^{-1}$. Between April 13 and April 20, 2009 the accelerator complex delivered a record of 73.1 pb$^{-1}$ per experiment in a single week and in March 2010 it delivered a record of 272.7 pb$^{-1}$ per experiment within a month. The Tevatron is expected to have delivered 11.1–12.1 fb$^{-1}$ per experiment in Run II by the end of September of 2011.

## LUMINOSITY IMPROVEMENTS

Run II started in the summer of 2001 and by the end of the year the instantaneous luminosity was in the range of (5-10) x 10$^{30}$ cm$^{-2}$s$^{-1}$. A lot of effort was invested in the follow up years in increasing the Tevatron luminosity (see Fig. 1 and 2). There is a long list of upgrades that took place, of which we list a few here. Understanding and tuning the Tevatron optics was a key contributor to the success. In 2002/2003 it was identified that the coupling in the Tevatron was not small, which led to significant emittance growth. During three long accelerator shudowns between 2003 and 2006 the global coupling around the machine was reduced by reshimming the Tevatron dipoles to address the coherent skew quadrupole component that was slowly growing. The completion of the Tevatron Beam Position Monitor (BPM) electronics upgrade in 2005 enabled more accurate beam lattice measurements, helped identify rolled quadrupoles, allowed for orbit stabilization within a store, and better monitoring of orbits store-to-store, resulting therefore in better reproducibility and enhanced reliability. In September 2005 a new Tevatron optics lattice was implemented decreasing the betatron amplitude function at the IPs, β*, from 35 cm to 29 cm. This increased the instantaneous luminosity by ~ 10%. At the same time the number of antiprotons available for Tevatron stores got increased by improvements in beam loading, longitudinal matching and damper optimization at the Main Injector, which resulted in an increase of protons at the antiproton production target by ~ 20%. In 2005 electron cooling at the Recycler became operational (maximum stash size of 6.08 x $10^{12}$ on March 21, 2011). In the mean time, since 2005 the antiproton stacking rate kept gradually improving with highest average stacking rate for one week of 25.65 x $10^{10}$/hr (January 2011). Additional electrostatic separators introduced in 2005/2006 allowed for an about 20% improvement in the luminosity lifetime, and implementation of the 1.7 GHz Schottky monitors and tune feedback improved it even further. The $2^{nd}$ order chromaticity compensation circuits implemented in 2007 allowed for higher proton intensity and also improved the luminosity lifetime. Tevatron alignment, which was performed almost every year during the long shutdowns, contributed significantly in the good performance of the machine. In 2009-2010 the shot setup duration was also reduced by about 20% allowing for faster turn-around between stores.

___________________________________________
∗ Work supported by DE-AC02-07CH11359
[#]vaia@fnal.gov

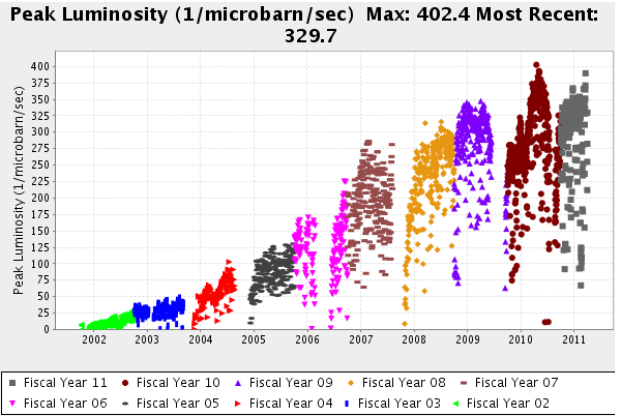

Figure 2: Tevatron peak luminosity in Run II (averaged over the CDF and D0 experiments) as a function of time.

Although there has been a lot of effort in increasing the instantaneous luminosity in Run II, the goal in the past few years has been to maximize the integrated luminosity to the experiments instead of trying to achieve higher instantaneous luminosities.

## LUMINOSITY MEASUREMENTS

### Machine

Absolute luminosity measurements by the machine are based on measurements of beam parameters like intensities, emittances, beam lattice, etc. and have an overall uncertainty of about 15-20%. In Eq. 1 one can see the dependence of the instantaneous luminosity on the number of protons and antiprotons per bunch, the emittances of the two beams and $\beta^*$. The emittances of the beams are directly related to the standard deviations of the beams spatially at the IPs. In Run II, the revolution frequency ($f_0$) is ~48 kHz and the number of bunches, $B$, is 36. $F(\sigma_z/\beta^*)$ is an hourglass form factor depending on the bunch length $\sigma_z$ and on $\beta^*$.

$$L = \frac{3\gamma f_0 N_p (BN_{\bar{p}})}{\beta^*(\varepsilon_p + \varepsilon_{\bar{p}})} F(\sigma_z/\beta^*) \qquad (1)$$

Intensities of protons and antiprotons are being measured by a Fast Bunch Integrator (FBI) of a Wall Current Monitor and by a Sampled Bunch Display (SBD). Both devices are ultimately scaled to a DC Current Transformer (DCCT) which is the most accurate device we have for measuring the total current in the Tevatron. The DCCT uncertainty is 1-2%. Typically, the measurement of the transverse beam emittances is based on the measurement of the transverse beam profiles. In the presence of dispersion, $D$, the momentum spread of the beam needs to be measured as well (see Eq. 2). Transverse beam profiles at the Tevatron are being measured by Flying Wire monitors, a synchrotron light monitor and ionization profile monitors. This allows for cross checking. The systematics of the Flying Wire (5 μm thick carbon fiber) measurements include the wire position measurement, the scintillator acceptance as a function of the beam position and the influence from previously scattered particles. The systematics of the synchrotron light monitor include the optical magnification, the non-uniformity and degradation of the intensifier as a function of time and the optical acceptance. The systematics of the Ionization Profile Monitors include resolution effects, the baseline subtraction, microchannel plate non-uniformities and degradation, etc.

$$\varepsilon_{tr} = \frac{6\pi(\beta\gamma)_{rel}}{\beta}\left[\sigma^2 - D^2\left(\frac{dp}{p}\right)^2\right] \qquad (2)$$

The momentum spread of the beam is being measured with the SBD. The overall uncertainty of emittance measurements is about 15%. In Fig. 3 we display Flying Wire normalized proton horizontal emittances through the accelerator chain as a function of time.

The beam optics parameters are being measured mainly within proton-only Tevatron stores using the orbit response method [1]. In Table 1 we show the $\beta^*$ and the dispersion at the CDF and D0 IPs as measured in the end of the long shutdown in 2010. The uncertainties vary between 5% (ideal case) and 15% depending on the goal of the measurement and coordination with other machine studies.

We routinely perform electrostatic separator scans to determine if the beam is well centered at the two IPs.

### CDF and D0 Experiments

The official Tevatron luminosity measurements are based on the relative luminosity measurements performed by the CDF and D0 experiments which are then normalized to a relatively well known and copious process, in this case the inclusive inelastic proton-antiproton cross section. The instantaneous luminosity $\mathcal{L}$ is being estimated by using Eq. 3, where μ is the average number of interactions per beam crossing, $f_{BC}$ is the frequency of bunch crossings, $\sigma_{in}$ is the inelastic cross section and $\varepsilon^{det}$ the acceptance of the detector. The average number of interactions can be estimated either by measuring the probability of zero interactions ( $P_0(\mu)=e^{-\mu}$ for a detector of 100% acceptance) or directly, by counting particles or hits or time clusters in the detector. The CDF and D0 collaborations have agreed to use a common proton-antiproton inelastic cross section for luminosity normalization in Run II. This common cross section has been derived [2] on the basis of averaging the inelastic cross sections measured by the Fermilab CDF and E811 experiments at 1.8 TeV and by extrapolating the cross section at 1.96 TeV.

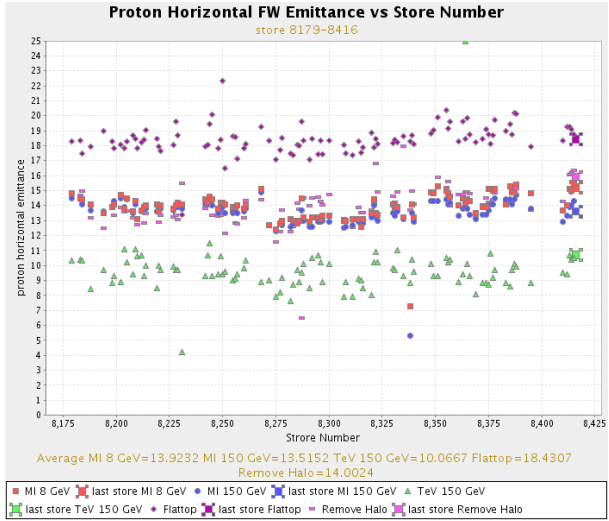

Figure 3: Normalized, 95% proton horizontal emittances through the accelerator chain as a function of Tevatron store number (time).

Table 1: Measured β* and D* parameters for the CDF and D0 interaction points in the end of August, 2010

|     | β*$_x$ proton | β*$_y$ proton | β* |
|-----|---------------|---------------|-----|
| CDF | 30.7 cm | 30.8 cm | 30.7 cm |
| D0  | 27.7 cm | 32.7 cm | 30.2 cm |
|     | D*$_x$ proton | D*$_y$ proton | D* |
| CDF | 1.1 cm | 1.7 cm | 2.0 cm |
| D0  | 1.4 cm | -0.7 cm | 1.6 cm |

$$\mu \cdot f_{BC} = \sigma_{in} \cdot \varepsilon^{det} \cdot \mathcal{L} \qquad (3)$$

In addition, luminosity measurements get cross calibrated with rarer, cleaner and better understood processes like the decay $W \to l \cdot \nu$.

Both experiments have used scintillating counters to measure the luminosity during Run I. For Run II where the instantaneous luminosity is substantially higher, CDF opted for a Cherenkov counter system while D0 for a scintillating system of better granularity than Run I.

The CDF Cherenkov counter system [3] consists of 48 counters per side arranged in 16 layers with 16 counters each covering the pseudorapidity region $3.7 \leq |\eta| \leq 4.7$. The counters are filled with isobutane and are being read by Hamamatsu R5800QCC Photomultipliers (PMTs) with quartz window. The Cherenkov counter system allows for good separation between primaries and secondaries, good amplitude resolution (~18% from photostatistics, light collection and PMT collection), good timing resolution and in addition it is radiation hard. In Fig. 4 is displayed the amplitude distribution for CDF data in one Cherenkov counter after an isolation requirement of less than 20 photoelectrons in the surrounding counters. The Single Particle Peak (SPP) is clear. Fig. 5 shows how the average number of particles (total amplitude over the amplitude of the SPP) or hits (counters with amplitude above a certain threshold) varies as a function of the average number of proton-antiproton interactions and compares the data with the Monte Carlo simulation. The data and the simulation compare very well. At the highest luminosities the particle counting algorithm is more linear. As a reference, note that μ approximately equal to 6 corresponds to $\mathcal{L}$ approximately equal to 2 x $10^{32}$ cm$^{-2}$s$^{-1}$. The CDF luminosity measurement is based as a default on measuring the probability of zero interactions and uses measuring hits and particles as a cross check. CDF has evaluated that the luminosity measurement using the probability of zero interactions is reliable up to about 3.6 x $10^{32}$ cm$^{-2}$s$^{-1}$. The current CDF luminosity measurement uncertainty is 5.8%. The leading contribution is from normalizing to the proton-antiproton inelastic cross section (4%). The next two most important contributions are due to simulating the material in the detector (3%) and the relative contribution from non-diffractive and diffractive processes in the Monte Carlo generator (2%). CDF is cross checking their absolute luminosity measurements by comparing with the inclusive W and Z boson cross section measurements and the comparison is very satisfactory. The yield of $J/\psi$'s and $W$'s through the $J/\psi \to \mu\mu$ and $W \to l\nu$ decays as a function of instantaneous luminosity serves as an additional check of the stability of the luminosity measurements. The aging rate of the PMTs is ~35% per fb$^{-1}$ and is being addressed by High Voltage and PMT gain adjustments or with replacements as needed.

The D0 Run II luminosity system [4] consists of two forward scintillator arrays covering the pseudorapidity region $2.7 \leq |\eta| \leq 4.4$. There are 24 wedges per array, each read out with a fine mesh PMT. Inelastic collisions are being identified by using the coincidence of in-time hits in the two arrays. Since October 2005 the luminosity readout electronics changed from NIM to custom VME [5]. The D0 luminosity measurement is based on measuring the probability of zero interactions. The current D0 luminosity measurement uncertainty is 6.1%. The leading contribution is from normalizing to the proton-antiproton inelastic cross section (4%). The next two most important contributions are due to the determination of the non-diffractive fraction (~4%) and the long term stability (~2.8%). Fig. 6 shows a data-Monte Carlo simulation comparison of counter multiplicity (above a threshold) assuming the final, non-diffractive fraction of 0.687±0.044. D0 is using the yield of forward muons as a function of time and instantaneous luminosity (see Fig. 7) as an additional check of the stability of the luminosity measurements (within ~1% during the past few years). The radiation damage to the scintillator is being addressed by annealing and replacement as needed.

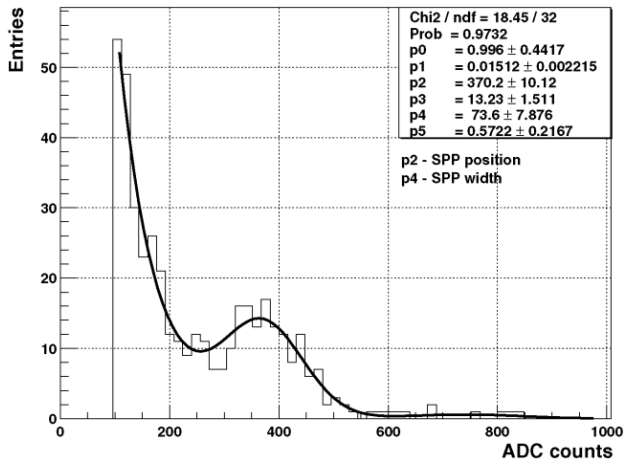

Figure 4: Amplitude distribution for a single Cherenkov counter at CDF. The solid line represents a fit to the data.

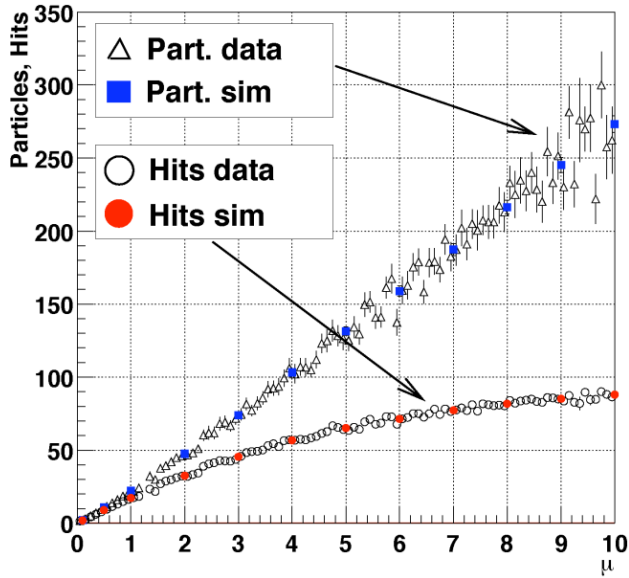

Figure 5: Data vs Monte Carlo simulation comparison of the average number of particles or hits vs the average number of proton-antiproton interactions at CDF.

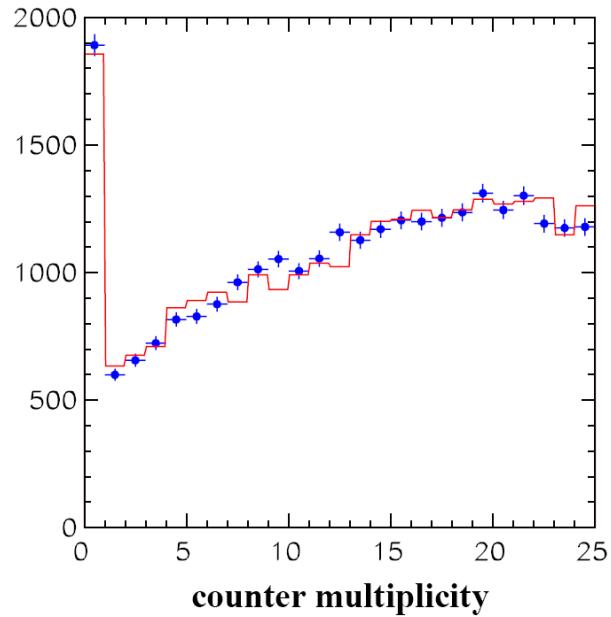

Figure 6: Data vs Monte Carlo simulation comparison of the multiplicity of the luminosity counters at D0 using the non-diffractive fraction. The points represent the data and the solid line the Monte Carlo. The plot corresponds to an instantaneous luminosity of $1.3 \times 10^{31}$ cm$^{-2}$s$^{-1}$.

*Luminosity Task Force*

In 2003 there was established the Luminosity Task Force, a joint effort between Accelerator, CDF and D0 colleagues, to address luminosity detector issues, beam position and beam width issues and other Tevatron issues affecting luminosity. We exchange information on a daily basis and meet as a larger group on a monthly basis or as needed. As a result, several machine studies have been performed and we have now a much better understanding of the Tevatron optics, the crossing angles and vacuum at the two IPs, the emittance of the proton and antiproton beams as well as of the luminosity detectors of both experiments. We monitor on a store-by-store basis luminosity related quantities for the experiments and the machine and examine their inter-correlations as well as possible correlations with external factors.

The CDF/D0 ratio of instantaneous luminosities is one of such quantities being checked on a store-by-store basis (see Fig. 8) and being compared with the expected ratio on the basis of beam parameters. The goal is to keep this ratio within a few percent around 1. Significantly larger deviations observed a few times so far have led to thorough investigation on both the machine and experiment sides and resulted to either machine parameter adjustments or to improvements in the techniques used by the experiments to measure the luminosity.

The CDF and D0 experiments provide as well on a regular basis measurements of beam parameters (beam position, beam emittance and β*) which are then being compared with the corresponding accelerator measurements. The CDF and D0 measurements are results of a fit of beam widths at the IPs as a function of z according to a model, where z is the axis along the direction of the collisions.

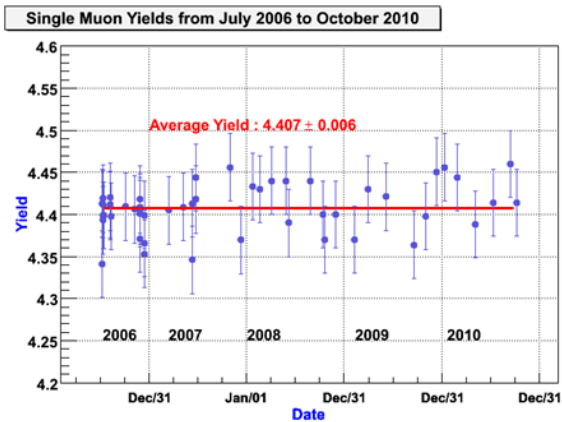

Figure 7: Single muon yields as a function of time, produced regularly by the D0 experiment to cross check the luminosity measurements.

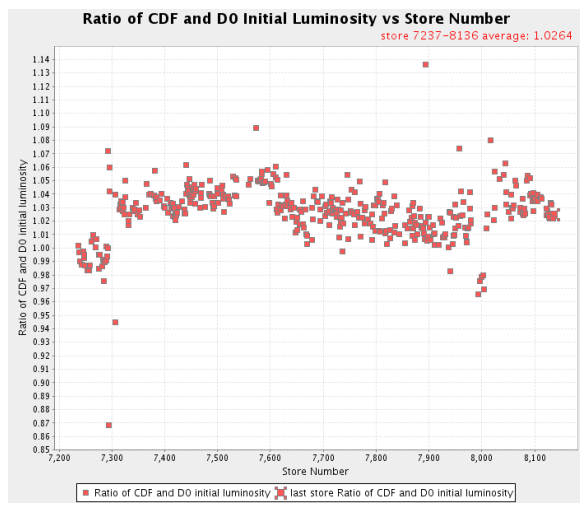

Figure 8: The ratio of the CDF and D0 initial luminosities in Tevatron stores as a function of time (store number) between October 2009 and September 2010.

## LESSONS LEARNED

Some of the lessons learned so far from the luminosity measurements at the Tevatron are: a) Continuous cross checking between the luminosities calculated by the machine and the luminosities measured by the experiments, as well as between the experiments themselves, is very valuable. The same is true for other beam parameters measured by the machine and the experiments; b) The method of counting zero interactions works well for the current Tevatron luminosities; c) Fine granularity detectors are needed for high instantaneous luminosities (Run I vs Run II); d) In situ calibration of the detector is very important; e) Detector stability is crucial; f) A good simulation of the processes involved and of the luminosity detector itself is needed as early as possible; g) A good knowledge of the physics cross section the measurement relies upon is necessary; h) Careful monitoring of gas purity when having a gas detector is a must; i) Minimizing - eliminating if possible - the dead time of the luminosity system is critical; j) Watchfulness is needed for aging due to large total luminosity and readinesss to replace consumables.

## CONCLUSION

Luminosity measurements at hadron-hadron colliders are very challenging. The luminosity uncertainty achieved at the Tevatron by the CDF and D0 experiments is approximately 6%, dominated by the uncertainty in the proton-antiproton inelastic cross section. The uncertainty achieved by the machine on the absolute luminosity measurement is approximately 15-20%. Cross checking the detector and machine luminosity measurements has been very valuable. We expect that the lessons learned from the Tevatron will be very useful for LHC.

I would like to thank the organizers for a very stimulating Workshop. I would also like to thank several colleagues from the Tevatron for discussions on the information presented here: N. Goldschmidt, B. Lee, R. Moore, A. Sukhanov, R. Tesarek, R. Thurman-Keup, A. Valishev and J. Zagel.